\documentclass[aps,prd,10pt,preprint,superscriptaddress,onecolumn,nofootinbib]{revtex4-1}
\usepackage{graphicx, epsfig} 
\usepackage{amsmath,amssymb,amsfonts,dsfont,mathrsfs,amsthm,mathtools}
\usepackage{bm} 
\usepackage{color}
\usepackage[usenames]{xcolor}
\usepackage{hyperref}
\usepackage{siunitx}
\hypersetup{linktocpage,colorlinks=true,urlcolor=blue,linkcolor=magenta,citecolor=blue}
\usepackage[normalem]{ulem}

\newcommand{\stkout}[1]{\ifmmode\text{\sout{\ensuremath{#1}}}\else\sout{#1}\fi}

\newcommand{\diff}[1]{\text{d}#1}

\newcommand{\Lie}{\mathcal{L}}
\newcommand{\Lag}{\mathscr{L}}

\begin{document}


\title{Topological Terms and the Misner String Entropy}


\author{Luca Ciambelli}
\email{ciambelli.luca@gmail.com}
\affiliation{Physique Th\'eorique et Math\'ematique, Universit\'e libre de Bruxelles and International Solvay Institutes, Campus Plaine C.P. 231, B-1050 Bruxelles, Belgium}

\author{Crist\'obal Corral}
\email{crcorral@unap.cl}
\affiliation{Instituto de Ciencias Exactas y Naturales, Facultad de Ciencias, Universidad Arturo Prat, Avenida Arturo Prat Chac\'on 2120, 1110939, Iquique, Chile}

\author{Jos\'e Figueroa}
\email{josepfigueroa@udec.cl}
\affiliation{Departamento de F\'isica, Universidad de Concepci\'on, Casilla, 160-C, Concepci\'on, Chile}

\author{Gast\'on Giribet}
\email{gaston@df.uba.ar}
\affiliation{Physics Department, University of Buenos Aires and IFIBA-CONICET, Ciudad Universitaria, Pabellón 1 (1428), Buenos Aires, Argentina}

\author{Rodrigo Olea}
\email{rodrigo.olea@unab.cl}
\affiliation{Universidad Andres Bello, Departamento de Ciencias F\'isicas, \\  Facultad de Ciencias Exactas, Sazi\'e 2212, Piso 7, Santiago, Chile}

\begin{abstract}
The method of topological renormalization in anti-de Sitter (AdS) gravity consists in adding to the action a topological term which renders it finite, defining at the same time a well-posed variational problem. Here, we use this prescription to work out the thermodynamics of asymptotically locally anti-de Sitter (AlAdS) spacetimes, focusing on the physical properties of the Misner strings of both the Taub-NUT-AdS and Taub-Bolt-AdS solutions. We compute the contribution of the Misner string to the entropy by treating on the same footing the AdS and AlAdS sectors. As topological renormalization is found to correctly account for the physical quantities in the parity-preserving sector of the theory, we then investigate the holographic consequences of adding also the Chern-Pontryagin topological invariant to the bulk action; in particular, we discuss the emergence of the parity-odd contribution in the boundary stress tensor.
\end{abstract}

\maketitle

\section{Introduction}

Hawking and Hunter argued in~\cite{Hawking:1998jf} that the existence of gravitational entropy is associated to a topological obstruction to foliation of the Euclidean section of the space, which results in an obstruction for the existence of a unitary Hamiltonian evolution. They were able to produce a universal formula meant to express the entropy of the spacetime in terms of defects that realize the obstructions to foliation. One example of such a defect is given by the Misner string~\cite{Misner:1963fr}, and so this seems to imply that the string does contribute to the entropy of the Taub-NUT spacetime~\cite{Taub:1950ez, Newman:1963yy}. In fact, the prime examples chosen in~\cite{Hawking:1998jf} to illustrate the phenomenon were precisely spaces with non-vanishing NUT charge, for which it was found that the entropy was not just a quarter the area, as it is for usual black holes~\cite{Bekenstein:1973ur}. This feature was further studied in~\cite{Hawking:1998ct}, where asymptotically locally anti-de Sitter (AlAdS) spacetimes were considered. For both Taub-Nut-AdS and Taub-Bolt-AdS solutions it was found that the additional contribution to the entropy coming from the Misner strings was essential to recover the entropy obtained from the usual Euclidean action calculation. This result for the entropy, on the other hand, seemed consistent with the AdS/CFT expectations. {Indeed,} soon later, Emparan, Johnson, and Myers reexamined the problem in AdS using holographic renormalization techniques~\cite{Emparan:1999pm}. The addition of boundary counterterms to the gravitational action renders the Euclidean action computation of AlAdS sacetimes finite and independent of the background subtraction. With this holography inspired method, finite quantities for Taub-NUT/Bolt-AdS spacetimes can be derived. 

The question as to whether gravitational entropy can be ascribed to spacetimes containing defects such as Misner strings, independently of the existence of event horizons, was also discussed in~\cite{Mann:1999pc, Mann:1999bt}, again using holographic renormalization {techniques}. By evaluating the entropy and Noether charges of Kerr-NUT/Bolt-AdS spacetimes, it was concluded in~\cite{Mann:1999bt} that, whenever the NUT charge is non-zero, the entropy does not equal one quarter of the area due to the contribution of the Misner string, in agreement with the observations of~\cite{Hawking:1998jf, Hawking:1998ct}.

In these twenty years, the geometry and thermodynamics of Taub-NUT/Bolt spacetimes have been studied by many authors and in many different contexts~\cite{Garfinkle:2000ms, Clarkson:2002uj, Clarkson:2003wa, Astefanesei:2004ji, Padmanabhan:2011ex, Johnson:2014xza, Clement:2015cxa, Araneda:2016iiy, Kubiznak:2019yiu, Bordo:2019tyh, Bordo:2019rhu, Donnay:2019zif, Durka:2019ajz, Kol:2020ucd, Porrati:2020anb, Kalamakis:2020aaj}.
Recently, special attention has been paid to the presence or absence of the Misner string~\cite{Johnson:2014xza, Kubiznak:2019yiu, Bordo:2019tyh, Bordo:2019rhu}. Here, we analyze the problem by using the method of topological renormalization; meaning, the method that consists of adding to the gravitational action a bulk piece of the Chern-Weil-Gauss-Bonnet (hereafter, Gauss-Bonnet) topological invariant, with the specific value of the coupling constant that renders the action equivalent to the MacDowell-Mansouri one~\cite{MacDowell:1977jt, Townsend:1977xw}. This procedure yields finite results for the Euclidean action, the Noether charges, and the thermodynamic quantities, while at the same time suffices to render the variational principle well-posed~\cite{Aros:1999id, Aros:1999kt}; see also~\cite{ Olea:2005gb, Olea:2006vd, Miskovic:2009bm, Giribet:2018hck}. The topological renormalization method has been shown to be consistent with holographic renormalization~\cite{Miskovic:2009bm,Anastasiou:2020zwc}, and so it will permit to analyze the problem in a way that is independent of the background substraction. In fact, we will show below that, while treating the AdS and AlAdS sectors in equal footing and in a unified scheme, we will obtain results for the thermodynamics of Taub-NUT/Bolt-AdS spacetimes that are in full agreement with the results found in the literature. In particular, we will analyze in detail the contribution to the entropy computation that comes from the Misner string. 

The use of topological invariants to support the original Hawking-Hunter argument makes more explicit the link between gravitational entropy and topological obstructions. It is thus natural to explore the effects of introducing other topological invariants and the consequences in the context of holography. The other topological invariant consisting in a dimension-four operator in four dimensions is the Chern-Pontryagin invariant, which carries opposite (odd) parity with respect to the Gauss-Bonnet one. The Chern-Pontryagin term does not affect the renormalization of the gravitational action, so its coupling constant is a priori arbitrary. The latter is reminiscent of the $\theta$-angle in gauge theories and can similarly be fixed by minimizing the value of the action when evaluated on self-dual solutions. We show that, adding this invariant to the bulk action, the boundary holographic stress tensor is augmented by the Hodge-dual Cotton tensor of the boundary. The full boundary stress tensor has now both even and odd parity contributions which realizes the so-called stress tensor/Cotton tensor duality~\cite{Bakas:2008gz, Mukhopadhyay:2013gja}. Combining the Cotton and stress tensor together has been instrumental in the fluid/gravity correspondence~\cite{Leigh:2011au, Leigh:2012jv, Caldarelli:2012cm, Petropoulos:2015fba, Ciambelli:2017wou}. This comes about thanks to the observation that the bulk Weyl tensor asymptotes to a specific combination of the latter~\cite{Leigh:2007wf, Mansi:2008br, Mansi:2008bs}. Since this combination is complex in Lorentzian signature, its parity odd piece, whose physical interpretation being challenging, has been dubbed ``reference tensor''; still, its contribution has shown to be important in the fluid/gravity fill-in problem~\cite{Gath:2015nxa}, where it controls the Petrov class in the bulk. Here, on top of boundary considerations, we explore the consequences of adding the bulk ancestor of the Cotton tensor --the Chern-Pontryagin term-- on the entropy.

The paper is organized as follows. In Sec.~\ref{II}, we review the fundamentals of Wald's formalism, which yields a procedure to compute the gravitational entropy. Section~\ref{III} is the bulk of the manuscript: we start showing the consequences on the Noether charge of adding the Gauss-Bonnet term to the bulk action, then we introduce the Taub-NUT/Bolt-AdS solutions and prove that the renormalized entropy and thermodynamics is recovered when the Gauss-Bonnet coupling constant is fixed to cancel divergences. Finally we explore the consequences of the  Chern-Pontryagin term both in the bulk thermodynamics and the boundary theory. We conclude offering some final comments and possible outlooks in Sec.~\ref{IV}.

\section{Noether-Wald formalism}\label{II}

In this section, we review the basics of Noether-Wald's formalism~\cite{Wald:1993nt,Iyer:1994ys} for the sake of completeness and to state conventions. Let us start by considering a general gravitational action principle defined on a $D$-dimensional manifold $\mathcal{M}$ provided with a metric $g_{\mu \nu }$ as 
\begin{align}\label{action}
 I\left[g_{\mu\nu}\right] = \int_{\mathcal{M}}\diff{^Dx}\sqrt{-g}\; \Lag\left[g_{\mu\nu},R^{\lambda\rho}{}_{\mu\nu} \right],
\end{align}
where $R^{\lambda\rho}{}_{\mu\nu} = g^{\rho\sigma}R^{\lambda}{}_{\sigma\mu\nu}$ is the Riemann tensor and $g=\det g_{\mu\nu}$. The metric is considered here as the only dynamical field: the connection is assumed to be torsion-free and metric compatible.\footnote{For extensions where these assumptions are relaxed see~\cite{Nester:1991yd,Chen:1998aw}.}

A stationary variation of the action~\eqref{action} yields~\cite{Wald:1993nt,Iyer:1994ys,Wald:1999wa,Padmanabhan:2011ex} 
\begin{align}\label{varaction}
 \delta I &= \delta \int_{\mathcal{M}}\diff{^Dx}\sqrt{-g}\,\Lag = \int_{\mathcal{M}}\diff{^Dx}\sqrt{-g}\,\delta g^{\mu\nu}\mathcal{E}_{\mu\nu} + \int_{\mathcal{M}}\diff{^Dx}\sqrt{-g}\,\nabla_\mu \Theta^\mu,
\end{align}
where
\begin{align}\label{generaleom}
 \mathcal{E}_{\mu\nu} &= E_{\mu}{}^{\lambda\rho\sigma}R_{\nu\lambda\rho\sigma} - \frac{1}{2}g_{\mu\nu}\Lag - 2 \nabla^\lambda\nabla^\rho E_{\mu\lambda\rho\nu}, \\
\label{generalbt}
 \Theta^\mu &= 2\delta\Gamma^{\lambda}{}_{\nu\rho} E_{\lambda}{}^{\rho\mu\nu} - 2\delta g_{\nu\sigma} \nabla_\rho E^{\nu\mu\rho\sigma} = -2\nabla_\rho\delta g_{\nu\sigma} E^{\rho\sigma\mu\nu} + 2\delta g_{\nu\sigma}\nabla_\rho E^{\rho\sigma\mu\nu}.  
 \end{align}
Here, $\mathcal{E}_{\mu\nu} = 0$ denotes the field equations and  we have defined the functional derivative of the Lagrangian with respect to the Riemannian curvature as\footnote{From hereon, we assume that $E_{\mu\nu\lambda\rho} = -E_{\nu\mu\lambda\rho}$, $E_{\mu\nu\lambda\rho} = - E_{\mu\nu\rho\lambda}$, $E_{\mu\nu\lambda\rho} = E_{\lambda\rho\mu\nu}$, and $E_{\mu[\nu\lambda\rho]}=0$.} 
\begin{align}
E_{\lambda\rho}{}^{\mu\nu} \equiv \frac{\partial\Lag}{\partial R^{\lambda\rho}{}_{\mu\nu}}.
\end{align}
This tensor plays a crucial role in Noether-Wald's formalism. The second term in Eq.~\eqref{varaction} is the boundary term arising from the variation of the action and it generically depends on the fields and variations thereof. By means of the Stokes' theorem, it can be expressed as
\begin{align}\notag
 \int_{\mathcal{M}}\diff{^Dx}\sqrt{-g}\;\nabla_\mu \Theta^\mu &= \int_{\partial\mathcal{M}}\diff{^{D-1}x}\sqrt{h}\; n_\mu \Theta^\mu \\
 \label{stokes}
 &= -2\int_{\partial\mathcal{M}}\diff{^{D-1}x} \sqrt{h}\;n_\mu\left[\nabla_\rho \delta g_{\nu\sigma} E^{\rho\sigma\mu\nu} - \delta g_{\nu\sigma}\nabla_\rho E^{\rho\sigma\mu\nu} \right],
\end{align}
where $h_{\mu\nu} = g_{\mu\nu} -  n_\mu n_\nu$ is the induced metric on the boundary, $h$ its determinant, and $n_\mu$ is a space-like unit normal such that $n_\mu n^\mu = 
1$ with $n^\mu h_{\mu\nu}=0$. 

Equation~\eqref{varaction} implies that diffeomorphism invariance leads to 
\begin{align}\label{conservationlaw}
 \nabla_\mu\left[\Theta^\mu(g,\Lie_\xi g) - \xi^\mu\Lag\right] \equiv \nabla_\mu J^\mu = -\Lie_\xi g^{\mu\nu}\mathcal{E}_{\mu\nu},
\end{align}
where $\Lie_\xi$ is the Lie derivative along the vector field $\xi$ and
\begin{align}
 J^\mu = -2\nabla_\nu\left(E^{\mu\nu\rho\sigma}\nabla_\rho\xi_\sigma + 2\xi_\rho\nabla_\sigma E^{\mu\nu\rho\sigma} \right).
\end{align}
The Noether current $J^\mu$ is conserved on-shell as it can be read off from Eq.~\eqref{conservationlaw}. The Poincaré lemma, in turn, implies that locally the Noether current $J^\mu$ can be written as 
\begin{align}\label{qmunu}
J^\mu = \nabla_{\nu}q^{\mu\nu} \;\;\;\;\; \mbox{where} \;\;\;\;\; q^{\mu\nu}= -2 \left(E^{\mu\nu\rho\sigma}\nabla_\rho\xi_\sigma + 2\xi_\rho\nabla_\sigma E^{\mu\nu\rho\sigma} \right) =  -q^{\nu\mu},
\end{align}
is known as the Noether prepotential. Thus, the conserved Noether charge associated to diffeomorphism invariance generated by the vector field $\xi$ is
\begin{align}\label{noethercharge}
 Q[\xi] = \frac{1}{2}\int_\Sigma\epsilon_{\mu_1\ldots\mu_D}q^{\mu_1\mu_2}\diff{x^{\mu_3}}\wedge\ldots\wedge\diff{x^{\mu_D}} \equiv \int_\Sigma Q_{\mu_1\ldots\mu_{D-2}}\diff{x^{\mu_1}}\wedge\ldots\wedge\diff{x^{\mu_{D-2}}} = \int_\Sigma\mathbf{Q} ,
\end{align}
where $\Sigma$ is a codimension-2 hypersurface. According to~\cite{Wald:1993nt}, the entropy is obtained when $\xi$ is an asymptotically time-like Killing vector and $\Sigma$ corresponds to the bifurcating horizon. In presence of additional obstructions to foliation, e.g. topological defects, other contributions to the entropy arise~\cite{Garfinkle:2000ms}. In general, it can be expressed as
\begin{align}\label{waldformula}
 S = \beta_\tau\int_{\Sigma}\mathbf{Q},
\end{align}
where $\beta_\tau$ is the period of the Euclidean time for the avoidance of conical singularities.

In the case of black holes in Einstein theory, Eq.~\eqref{waldformula} implies that the entropy is one quarter of the horizon's area. However, a remarkable counterexample appears in Taub-NUT/Bolt geometry with spherical base manifold, since the presence of Misner strings breaks down the standard entropy/area relation~\cite{Hawking:1998ct,Emparan:1999pm,Mann:1999pc,Garfinkle:2000ms,Astefanesei:2004ji}. The modification of the entropy law stems from the obstruction to foliate the spacetime with constant-time hypersurfaces, turning the contribution from the Misner string nontrivial at the poles. Through the Noether-Wald formalism, this was computed in Ref.~\cite{Garfinkle:2000ms} for asymptotically locally flat (AlF) spaces. In AdS, in contrast, the entropy of the Misner string defined by Wald's formulae becomes divergent. In the following, we show that this problem can be circumvented by introducing the Gauss-Bonnet invariant into the gravitational action.

\section{Topological renormalization}\label{III}

In four dimensions, the Einstein-Hilbert action in presence of the Gauss-Bonnet term is\footnote{We call this the Einstein-Gauss-Bonnet action and its integrand the Einstein-Gauss-Bonnet Lagrangian.}
\begin{align}\label{EGB}
 I_{\rm EGB}[g_{\mu\nu}] = \kappa\int\diff{^4x}\sqrt{-g}\left(R-2\Lambda + \alpha\mathcal{G}  \right),
\end{align}
where $\kappa= \left(16\pi G\right)^{-1}$ is the gravitational constant, $\alpha$ is the Gauss-Bonnet coupling constant and
\begin{align}
\label{GB}
 \mathcal{G} &= R^2 - 4R_{\mu\nu}R^{\mu\nu} + R_{\mu\nu\lambda\rho}R^{\mu\nu\lambda\rho} = 3!\delta_{[\mu}^{[\alpha}\delta_\nu^\beta\delta_\lambda^\gamma\delta_{\rho]}^{\delta]} R^{\mu\nu}{}_{\alpha\beta} R^{\lambda\rho}{}_{\gamma\delta}.
\end{align}

The field equations obtained from the action~\eqref{EGB} and the off-shell functional derivative of the Einstein-Gauss-Bonnet Lagrangian with respect to the Riemann tensor are 
\begin{align}\label{eomEGB}
 \mathcal{E}_{\mu\nu} &= R_{\mu\nu} - \frac{1}{2}g_{\mu\nu}R + \Lambda g_{\mu\nu} = 0,\\
 \label{EEGB}
 E^{\mu\nu}{}_{\lambda\rho} &= \kappa\left(\delta_{[\lambda}^{[\mu}\delta_{\rho]}^{\nu]} + 12\alpha\delta_{[\lambda}^{[\mu}\delta_\rho^\nu\delta_\alpha^\gamma\delta_{\beta]}^{\delta]} R^{\alpha\beta}{}_{\gamma\delta}\right),
\end{align}
respectively. Equation~\eqref{eomEGB} is solved by Einstein spaces with $R_{\mu\nu} = \Lambda g_{\mu\nu}$. Thus, on-shell, the Riemann tensor can be written as 
\begin{align}
 R^{\mu\nu}{}_{\lambda\rho} &= W^{\mu\nu}{}_{\lambda\rho} - \frac{2}{\ell^2}\delta_{[\lambda}^{[\mu}\delta_{\rho]}^{\nu]},
\end{align}
where $\Lambda=-3/\ell^2$ and the Weyl tensor has been defined as
\begin{equation}
W_{\mu\nu\lambda\rho }=R_{\mu\nu\lambda\rho }+\frac 12 \Big(R_{\mu \rho}g_{\nu\lambda}-R_{\nu\rho}g_{\mu\lambda}+R_{\nu\lambda}g_{\mu \rho}-R_{\mu\lambda}g_{\nu\rho}\Big)+\frac 16 R\Big(g_{\mu\lambda}g_{\nu\rho}-g_{\mu\rho}g_{\nu\lambda}\Big).
\end{equation}
For Einstein-AdS spaces, Eq.~\eqref{EEGB} becomes
\begin{align}\label{EEGBonshell}
 E^{\mu\nu}{}_{\lambda\rho} &= \kappa\left(1 - \frac{4\alpha}{\ell^2} \right)\delta_{[\lambda}^{[\mu}\delta_{\rho]}^{\nu]} + 2\kappa\alpha W^{\mu\nu}{}_{\lambda\rho},
\end{align}
and, therefore, the Noether prepotential reads 
\begin{align}\label{qmunuEGB}
q^{\mu\nu} = -2\kappa\left[\left(1-\frac{4\alpha}{\ell^2}\right)\delta_{[\lambda}^{[\mu}\delta_{\rho]}^{\nu]} + 2\alpha W^{\mu\nu}{}_{\lambda\rho}  \right]  \nabla^\lambda\xi^\rho.  
\end{align}

It is well known that there exists a proper choice of the Gauss-Bonnet coupling in four dimensions renormalizing the Noether charge~\cite{Aros:1999id} and Euclidean on-shell action~\cite{Olea:2005gb} for AlAdS solutions. Its particular value is obtained by demanding that the first term on the right-hand side of Eq.~\eqref{EEGBonshell} vanishes, which is justified by the fact the Weyl tensor is the only combination between the curvature and the metric that has the correct falloff.\footnote{The appearance of the Weyl tensor at the boundary is
the key ingredient to make contact with the notion of Conformal Mass 
in AAdS spacetimes \cite{Ashtekar:1984zz,Jatkar:2014npa}.}

The very same choice allows one to renormalize the entropy of Misner string in Euclidean Taub-NUT/Bolt-AdS solutions. This is remarkable as one could have expected that the coefficient that suffices to renormalize solutions in the asymptotically AdS sector would differ from the one that does the job in other AlAdS sectors. More concretely, since different values of the NUT charge define different asymptotic sectors, one could have expected that the coupling of the topological term would in general depend on that charge, but notably it does not. Here, we will show that a similar phenomenon occurs in the thermodynamics calculation, where the integration is not performed at infinity.

\subsection{Taub-NUT/Bolt-AdS}

The field equations~\eqref{eomEGB} are solved by the Euclidean inhomogeneous stationary metric constructed on complex line bundles over $S^2$, that is
\begin{align}\label{ElTaubNUTgeneral}
 \diff{s^2} &= f(r)\left(\diff{\tau} + 2n\cos\theta \, \diff{\phi} \right)^2 + \frac{\diff{r^2}}{f(r)} + (r^2-n^2)\left(\diff{\theta^2} + \sin^2\theta \, \diff{\phi^2} \right),
\end{align}
where
\begin{align}\label{fsol}
 f(r) = \frac{r^2+n^2}{r^2-n^2} - \frac{2MGr}{r^2-n^2} + \frac{r^4 - 6r^2n^2 - 3n^4}{\ell^2\left(r^2-n^2 \right)},
\end{align}
and $\Lambda = -3/\ell^2$. This is the asymptotically AdS version~\cite{Hawking:1998ct} of the Taub-NUT spacetime~\cite{Taub:1950ez, Newman:1963yy}, whose Euclidean version can be regarded as a gravitational instanton~\cite{Eguchi:1978xp, Gibbons:1979zt}. Here, $M$ is an integration constant associated to the mass, while $n$ is a constant that can be thought of as the gravitational magnetic charge $N=n/G$. Nuts and bolts, alongside the absence of conical singularities, are characterized by
\begin{align}
 &\mbox{For NUT:} & f(n) &= 0& &\mbox{and}& f'(r)\big|_{r=n} &= \frac{4\pi}{\beta_\tau}, \\
 &\mbox{For Bolt:} & f(r_b) &= 0& &\mbox{and}& f'(r)\big|_{r=r_b} &= \frac{4\pi}{\beta_\tau},
\end{align}
where $r_b>n$ is the bolt radius. These conditions determine the set of fixed points to be one- and two-dimensional, respectively. In turn, they fix the integration constant $M$ as
\begin{align}\label{mnut}
 M &= \frac{n}{G}\left(1- \frac{4n^2}{\ell^2} \right) \equiv M_{\rm NUT} ,\\
 M &= \frac{1}{2Gr_b}\left(r_b^2+n^2 + \frac{r_b^4 - 6r_b^2n^2 - 3n^4}{\ell^2} \right) \equiv M_{\rm Bolt} ,\label{MBolt}
\end{align}
respectively. For NUT, the period of the Euclidean time is $\beta_\tau=8\pi n$, and the Weyl tensor is globally self-dual. That implies that the total mass, in terms of the electric and magnetic mass of the solution is identically zero~\cite{Araneda:2016iiy}. Therefore, for different values of the NUT charge, this solution represents topologically inequivalent vacuum states. For bolt, on the other hand, the Weyl tensor is not globally self-dual and the period of the Euclidean time is given by
\begin{align}
 \beta_\tau &= \frac{4\pi r_b}{1 + \frac{3}{\ell^2}\left(r_b^2-n^2 \right)}.\label{betaBolt}
\end{align}
Additionally, the requirement of the Misner string being unobservable\footnote{By unobservable we mean that bolt solutions should have the same temperature as NUT ones. As we will see, they still have different entropy though.} \cite{Misner:1963fr} imposes a further condition, i.e. $\beta_\tau=8\pi n$, which relates the bolt radius with the NUT charge according to
\begin{align}\label{rb}
 r_b &= \frac{\ell^2}{12 n}\left[1\pm\sqrt{1 - \frac{48n^2}{\ell^2}\left(1-\frac{3n^2}{\ell^2} \right)} \right].
\end{align}
Reality and positivity of the bolt radius impose a range on the NUT charge such that the solution exists.

The role of the Gauss-Bonnet term is crucial to renormalize the entropy contribution of the Misner string. The relevant components of the Noether charge~\eqref{noethercharge} are computed by inserting the off-shell value of~\eqref{EEGB} into Eq.~\eqref{qmunu}; they read
\begin{align}\notag
 Q_{\theta\phi} &= \kappa\left(r^2-n^2 \right)f'\sin\theta \\
 & + \frac{4\kappa\alpha}{\left(r^2-n^2\right)^2}\left[\left(r^2-n^2 \right)^2 f' - \left(r^4+4r^2n^2-5n^4 \right)f' f + 4f^2 r n^2 \right]\sin\theta, \\
 Q_{r\phi} &= -\frac{4\kappa n^2 f\cos\theta}{r^2-n^2} + \frac{8\kappa\alpha n^2}{\left(r^2-n^2 \right)^2}\left[\left(r^2-n^2 \right)\left(f'' f + f'^2 \right) - 2f'fr \right]\cos\theta.
\end{align}
Thus, the Noether charge in this case is
\begin{align}\notag
\int_\Sigma\mathbf{Q} &= \int_{r=r_b}\mathbf{Q} + \int_{\theta=0}\mathbf{Q} + \int_{\theta=\pi}\mathbf{Q} \\
\label{noethereval}
&= \int_0^{2\pi}\diff{\phi}\int_0^\pi\diff{\theta}\; Q_{\theta\phi}\big|_{r=r_b} - \int_0^{2\pi}\diff{\phi}\int_{r_b}^\infty\diff{r}\; Q_{r\phi}\big|_{\theta=0} +  \int_0^{2\pi}\diff{\phi}\int_{r_b}^\infty\diff{r}\; Q_{r\phi}\big|_{\theta=\pi}\,,
\end{align}
where the last two terms arise from the contribution of the Misner string. In absence of the cosmological constant, these integrals were computed in Ref.~\cite{Garfinkle:2000ms} giving $S=4\pi n^2$ and $S=5\pi n^2$ for nuts and bolts, respectively. Their difference yields to the same result as in Ref.~\cite{Hawking:1998ct} without infinite background subtraction. In presence of the cosmological constant, however, the last two terms are divergent. One possibility to circumvent this problem is to add proper counterterms and employ Euclidean methods as in Ref.~\cite{Mann:1999pc}. The same procedure leads to the aforementioned entropies in absence of the cosmological constant.

Here, we notice that an alternative method to renormalize the contribution of the Misner string in AdS is to fix the Gauss-Bonnet coupling as
\begin{align}\label{alphasol}
 \alpha = \frac{\ell^2}{4}.
\end{align}
With this choice, the action~\eqref{EGB} becomes the one of  MacDowell-Mansouri\footnote{In five dimensions the quadratic Gauss-Bonnet term does change the theory at classical level, and for the same choice of couplings constants (\ref{alphasol}) the action of the gravitational theory coincides with that of five-dimensional Chern-Simons; see \cite{Zanelli:2005sa} and references therein.}~\cite{MacDowell:1977jt, Townsend:1977xw} and, on-shell, it can be written as the conformal invariant contribution~\cite{Araneda:2016iiy}
\begin{equation}\label{acWW}
I_{\text{EGB}}=\frac{\kappa\ell^2}{ 4}\int \diff{^4 x} \sqrt{-g}\, W_{\mu\nu\lambda\rho}W^{\mu\nu\lambda\rho }.
\end{equation}
Additionally, the Noether prepotential for Einstein spaces becomes
\begin{align}\label{qmunuEGBonshell}
 q^{\mu\nu} = -\kappa\ell^2\, W^{\mu\nu}{}_{\lambda\rho} \nabla^\lambda\xi^\rho.
\end{align}

Thus, Eq.~\eqref{waldformula} yields the renormalized entropy
\begin{align}\label{Sbolt}
 S_{\rm Bolt} &= \frac{32\pi^2 \kappa n}{r_b}\left[4r_b^2 - 2n^2 + \frac{1}{\ell^2}\left(3r_b^4 - 12r_b^2n^2 - 3n^4 + \ell^4 \right) \right],
\end{align}
which, after replacing (\ref{rb}), can be seen to agree with the result in the literature \cite{Johnson:2014xza} up to a (thermodynamic irrelevant) constant piece $\Delta S= \pi\ell^2/G$. It is worth noticing that we could have done the computation in a different gauge by considering in (\ref{ElTaubNUTgeneral}) the Misner change of coordinate $\tau \to \tau \pm 2n\phi $, which suffices to eliminate one of the Misner strings, namely the one located at $\theta= \pi/2(1\pm 1)$, at the price of introducing closed timelike curves in the Lorentzian geometry. As expected, the result of the computation performed in this way agrees with (\ref{Sbolt}), with one of the last terms in the second line of (\ref{noethereval}) contributing zero in that case, but being it compensated by the other. 

The advantage of using Noether-Wald procedure for the $4D$ Einstein-Gauss-Bonnet action is that the resulting formula is fully covariant. Thus, it properly adapts to include all the boundaries present in the geometry. In addition, it can systematically be generalized to higher, even dimensions.  

The lowest order in the saddle-point approximation from the quantum statistical relation $\ln Z \approx - I$, where $Z$ is the partition function and $I$ is the on-shell Euclidean action, provides a correct thermodynamic description of gravitational solutions. Remarkably, the same result for the entropy can be obtained by choosing the Gauss-Bonnet coupling in Eq.~\eqref{alphasol}, such that the renormalized  Euclidean action is 
\begin{align}\label{Ibolt}
 I_{\rm Bolt} = -\frac{32\pi^2\kappa n}{r_b}\left[2r_b^2 - 4n^2 + \frac{1}{\ell^2}\left(r_b^4+3n^4 + \ell^4 \right) \right].
\end{align}
It is straightforward to check that this expression yields to the same entropy obtained from the Wald's formalism [cf. Eq.~\eqref{Sbolt}] through
\begin{align}
 S_{\rm Bolt}=\beta_\tau\frac{\partial I_{\rm Bolt}}{\partial\beta_\tau} - I_{\rm Bolt}.\label{KLKL}
\end{align}
Thus, the Gibbs-Duhem relation $S=\beta_\tau H_\infty - I$ allows one to obtain the energy of the system as
\begin{align}\label{lamasita}
 H_\infty &= \frac{8\pi\kappa}{r_b}\left[r_b^2+n^2 + \frac{1}{\ell^2}\left(r_b^4-6r_b^2n^2-3n^4 \right) \right],
\end{align}
which is exactly the same as $M$ in Eq.~\eqref{MBolt}.

Notice that, when comparing with the results in~\cite{Johnson:2014xza}, the entropy~\eqref{Sbolt} has an additional contribution $\Delta S= \pi\ell^2/G$. In the case of topological black holes, this shift in the entropy is proportional to the Euler characteristic of the horizon~\cite{Olea:2005gb}; i.e. it computes a topological number of the constant-$\tau $ slices of the horizon. As $\Delta S$ is just an additive constant, it does not affect relevant thermodynamic relations such as the first law. On the other hand, the value of~\eqref{Ibolt} differs from those in \cite{Johnson:2014xza, Emparan:1999pm} by a term $\Delta I=-\Delta S$ so that (\ref{KLKL}) holds, showing that $\Delta S$ indeed reflects the horizon topology. For $r_b=n$, the result (\ref{Sbolt}) takes the form
\begin{align}\label{SboltNUT}
S_{\rm NUT} &= \frac{4\pi n^2}{G} \left(1  - \frac{6n^2}{\ell^2} \right)+\frac{2\pi \ell^2 }{G},
\end{align}
which also agrees with the entropy of the Taub-NUT-AdS solution given in \cite{Mann:1999pc, Emparan:1999pm, Johnson:2014xza} up to a constant, cf. \cite{Hawking:1998ct}.\footnote{Our notation relates to that of \cite{Hawking:1998ct} by identifying parameters as $\ell=2b , n=b\sqrt{E}, k=1, s=r_b/(b\sqrt{E})$ and rescaling the time and radial coordinates by $2b$ and $b\sqrt{E}$, respectively.}

\subsection{Adding the Chern-Pontryagin invariant}

%
Having shown above that the addition of the Gauss-Bonnet invariant suffices to renormalize the gravitational action and yields the correct result for the charges, it is natural to ask about the effects of including the only other dimension-four topological operator in four dimensions, i.e. the Chern-Pontryagin topological invariant. The latter reads
\begin{equation}
{\cal P}_4=-\frac{1}{2} \star R^{\mu\nu}_{\ \rho \eta }R_{\ \mu\nu}^{\rho\eta }=-\frac{1}{4} \varepsilon^{\mu\nu }_{\ \alpha \beta }R^{\alpha \beta }_{\ \rho \eta }R^{\rho \eta }_{\ \mu\nu },\label{P4}
\end{equation}
where we defined the left Hodge dual of the Riemann tensor following the convention $\star R_{\mu\nu\rho\eta }\equiv \tfrac{1}{2}\varepsilon_{\mu \nu \alpha \beta }R^{\alpha \beta }{}_{\rho \eta }=\frac{1}{2}\varepsilon_{\mu \nu \alpha \beta }R^{\alpha \beta }_{\ \rho \eta }$. Evaluated on the Taub-NUT-AdS geometry this yields
\begin{equation}
-\frac{1}{8\pi^2}\int d^4x\,\sqrt{-g} \, {\cal P}_4=\frac{1}{32\pi^2} \int d^4x\,\sqrt{-g}\, \varepsilon^{\mu\nu }_{\ \alpha \beta }R^{\alpha \beta }_{\ \rho \eta }R^{\rho \eta }_{\ \mu\nu }=2-\frac{16n^2}{\ell^2}\left(1-\frac{2n^2}{\ell^2}\right)\, .\label{signatura}
\end{equation}
This value, which reduces to the well-known result $2$ for the Ricci flat metric in the limit $\ell \to \infty$, changes its global sign if one perform the change $n\to -n$ in (\ref{ElTaubNUTgeneral}) (see Ref.~\cite{Hawking:1976jb}). We also see from this expression that something special occurs for $n=\pm \ell/2$, for which (\ref{signatura}) vanishes. Notice that the mass (\ref{mnut}) is also zero for such especial values of $n$. As discussed in \cite{Johnson:2014xza}, the geometries with $n=\pm \ell/2$ correspond to AdS$_4$ spacetime with a non-trivial slicing, the trivial $S^2\times S^1$ slicing corresponding to $n=M=0$. It is worth noticing that for $n=\pm \ell/2$ the entropy formula (\ref{SboltNUT}) gives a positive result, avoiding the puzzle of negative entropy observed in \cite{Chamblin:1998pz, Mann:1999pc, Johnson:2014xza}. Notice also that for $n=\pm \ell/\sqrt{2}$ the value of the signature (\ref{signatura}) reduces to that of Ricci flat Taub-NUT. It is the dimensionless ratio $n / \ell $ what controls the value of the Pontryagin invariant, so suggesting that the values that this ratio takes might be important for the quantum theory. 

The Chern-Pontryagin term \eqref{P4} has opposite parity than the Gauss-Bonnet one, such that its inclusion results in an action that is not longer parity even.
As the ${\cal P}_4$ can be written as
\begin{equation}
{\cal P}_4=-\frac{1}{2}\star W^{\mu\nu }_{\ \rho \eta } W^{\rho \eta }_{\ \mu \nu }, \label{WP4}
\end{equation}
when adding it to the gravitational Lagrangian one finds the on-shell renormalized action
\begin{equation}
I_{\text{EGBP}}[g_{\mu\nu }]=\frac{\kappa\ell^2}{4}\int d^4 x \sqrt{-g}\, W^{\mu\nu }_{\ \rho\eta }W^{\rho \eta }_{\ \mu\nu }+\vartheta\, \frac{\kappa \ell^2}{ 4}\int d^4 x\sqrt{-g} \, \star W^{\mu\nu }_{\ \rho \eta } W^{\rho \eta }_{\ \mu \nu},\label{GHGG}
\end{equation}
where $\vartheta $ is a constant that is reminiscent of the coupling of the $\theta$-term in gauge theory. In fact, the action (\ref{GHGG}) can be thought of as the MacDowell-Mansouri action augmented with a $CP$ violating term. In contrast to what happens with the coupling $\alpha $ of the Gauss-Bonnet invariant, which can be fixed as in~\eqref{alphasol} by demanding regularity of the action, the value of $\vartheta $ is not fixed by such a requirement. Therefore, it is necessary to apply a different criterion. A natural one would be asking the action to vanish when evaluated on the self-dual vacuum solution. This results in the value\footnote{The two signs here correspond to the self-dual and the anti-self-dual solutions.}
\begin{equation}\label{tita}
\vartheta = \pm 1 \, .
\end{equation}
Using this, the on-shell action can be written as~\cite{Araneda:2016iiy}
\begin{equation}\label{actEGBPonshell}
I_{\text{EGBP}}=\frac{\kappa\ell^2}{32}\int_{\mathcal{M}}\sqrt{-g}\;\delta_{[\mu}^{[\alpha}\delta_\nu^\beta\delta_\lambda^\sigma\delta_{\rho]}^{\tau]} \left(W^{\mu\nu}{}_{\alpha\beta} \pm \star W^{\mu\nu}{}_{\alpha\beta} \right)\left(W^{\lambda\rho}{}_{\sigma\tau} \pm \star W^{\lambda\rho}{}_{\sigma\tau} \right). 
\end{equation}
Since the Taub-NUT-AdS solution with $M$ fixed by Eq.~\eqref{mnut} is globally (anti-)self dual, the action~\eqref{actEGBPonshell} vanishes identically when evaluated at this configuration. For bolt, we take the case $n>0$ for the sake of simplicity. Thus, the renormalized on-shell action in the presence of the Chern-Pontryagin term with fixed theta parameter is
\begin{align}\label{Iboltpont}
 I_{\rm Bolt} = -\frac{16\pi^2\kappa n \left(2r_b + n \right)}{r_b^2}\left[2\left(r_b-n \right)\left(r_b + 3n \right) +\frac{1}{\ell^2}\left[\left(r_b-n \right)^2\left(r_b+3n \right)^2 + \ell^4 \right] \right],
\end{align}
which, in contrast to (\ref{Ibolt}), for $r_b=n$ is a constant independent on $n$. This is expected: since the action~\eqref{actEGBPonshell} vanishes when evaluated at the Taub-NUT-AdS solution, Eq.~\eqref{Iboltpont} can differ from zero only by a $n$-independent constant when $r_b=n$.

Additionally, the Noether's prepotential for Einstein-AdS spaces becomes
\begin{align}\label{qmunuEGBPonshell}
 q^{\mu\nu} = -\kappa\ell^2\left(W^{\mu\nu}{}_{\lambda\rho} \pm \star W^{\mu\nu}{}_{\lambda\rho} \right)\nabla^\lambda\xi^\rho.
\end{align}
Therefore, the Noether charge vanishes identically for (anti-)self dual solutions as well. This, in turn, implies that the total mass and entropy are zero for Taub-NUT-AdS. In other words, the role of the Chern-Pontryagin term with (\ref{tita}) is to set Taub-NUT-AdS solution as the background reference. 


The entropy for bolt obtained through the Wald's formula in  Eq.~\eqref{waldformula} via the Noether's prepotential~\eqref{qmunuEGBPonshell} yields
\begin{align}\notag
 S_{\rm Bolt} &= \frac{16\pi^2\kappa n}{r_b^2}\Big[2\left(r_b - n \right)\left(4r_b^2 + 5r_b n + 3n^2 \right) \\ 
 \label{Sbolpont}
 &+ \frac{1}{\ell^2}\left[3\left(r_b+3n \right)\left(r_b-n \right)^2\left(2r_b^2+nr_b + n^2\right) + \ell^4\left(2r_b+n \right) \right] \Big],
\end{align}
which, in contrast to (\ref{Sbolt})-(\ref{SboltNUT}), when $r_b=n$ is independent on $n$. 

It is straightforward to check that the entropy~\eqref{Sbolpont} can be obtained from the thermodynamic relation~\eqref{KLKL} using the renormalized action~\eqref{Iboltpont}. The thermodynamic mass for bolt, on the other hand, is obtained from standard methods, yielding 
\begin{align}\label{LAMasa}
 \hat{M}_{\rm Bolt} &= \frac{\partial I_{\rm Bolt}}{\partial \beta_\tau} =  \frac{8\pi\kappa\left(r_b -n \right)^2}{r_b}\left[ 1 + \frac{\left(r_b-n \right)\left(r_b+3n \right)}{\ell^2}   \right] = M_{\rm Bolt} - M_{\rm NUT},
\end{align}
where $M_{\rm Bolt}$ and $M_{\rm NUT}$ are the charges given by (\ref{lamasita}). The mass (\ref{LAMasa}) is equivalent to the definition of Ref.~\cite{Araneda:2016iiy} in terms of the Noether charge when $\vartheta=+1$.

\subsection{Boundary stress tensors}

The topological renormalization method is consistent with holographic renormalization \cite{Miskovic:2009bm}, leading to reproduce the boundary stress tensor. The inclusion of the Chern-Pontryagin term in the gravity action on AlAdS spaces modifies the result one obtains by means of holographic renormalization for the dual theory stress tensor $T_{ab}$. In fact, for $\vartheta = \pm 1$ one obtains\footnote{The coefficient of the second term in (\ref{Tpm}) acquires an imaginary $i$ factor in the Lorentzian case, due to the presence of the Levi-Civita pseudotensor in the definition of $C_{ab}$, Eq. \eqref{Cotton}. Our conventions on the Wick rotation from Euclidean to Lorentzian are $n\to i n$, $\tau\to i \tau$ and $C_{ab}\to -i C_{ab}$.}
\begin{equation}
T^\pm_{ab}=T_{ab}\mp \frac{\ell^2}{8\pi G}C_{ab}\, , \label{Tpm}
\end{equation}
where $a,b=0,1,2$, $G$ is the four-dimensional Newton constant and we introduced the Cotton tensor
\begin{equation}\label{Cotton}
    C_{ab}=\sqrt{-g^{(0)}}\varepsilon_{a}{}^{cd}\nabla_c\left(R^{(0)}_{db}-\frac{1}{4}R^{(0)}g^{(0)}_{db}\right).
\end{equation}
Here $g^{(0)}_{ab}$ is the boundary metric while $R^{(0)}_{ab}$ and $R^{(0)}$ its Ricci tensor and scalar curvature, respectively. The tensor $T^{\pm}_{ab}$ is the full stress tensor, which consists of the sum of the holographic part of standard Balasubramanian-Kraus tensor $T_{ab}$ and the Cotton tensor $C_{ab}$, with a specific relative coefficient.\footnote{The coupling of tensor (\ref{Tpm}) would result in a boundary model of Topologically Massive Gravity \cite{Deser:1982vy}.} This is related to fluid interpretation of the dual stress tensor (named ``reference tensor'') as well as to the so-called stress tensor / Cotton tensor duality, cf. \cite{Leigh:2007wf, Mansi:2008br, Mansi:2008bs, Bakas:2008gz, Leigh:2011au, Leigh:2012jv, Miskovic:2009bm}. We will not discuss this in extenso here as it lies beyond the scope of the present paper. Here, we focused on the AlAdS sector; more precisely, on the thermodynamics of the Misner strings of both Taub-NUT-AdS and Taub-Bolt-AdS solutions as it is described by the topological renormalization method. 

The Cotton tensor in (\ref{Tpm}) is the parity odd part of $T^\pm_{ab}$, while the Balasubramanian-Kraus piece is parity even. This implies that a parity transformation results in an interchange $T^+_{ab}\leftrightarrow T^-_{ab}$. We discuss the parity transformation below, after writing the explicit expressions for the tensors; see (\ref{ELEstreso})-(\ref{ELCotton}). Let us just mention here that the presence of the NUT charge $n$ in the boundary metric, being this charge parity odd, makes natural to consider in the boundary theory both components in (\ref{Tpm}). 

Let us now evaluate the full stress-tensor $T^\pm_{ab}$ on the Taub-NUT/Bolt-AdS metrics. To do that, it is convenient to rewrite the metric (\ref{ElTaubNUTgeneral}) by performing the Misner change of coordinate $\tau \to \tau - 2n \phi $ followed by the inversion $\phi \to -\phi$. This yields
\begin{align}\label{ElTaubNUTgeneralrewritten}
 \diff{s^2} &= f(r)(\diff{\tau} + 4n\sin^2({\theta}/{2})\, \diff{\phi} )^2 + \frac{\diff{r^2}}{f(r)} + (r^2-n^2)\left(\diff{\theta^2} + \sin^2\theta\, \diff{\phi^2} \right).
\end{align}
Evaluating the Balasubramanian-Kraus boundary tensor on this metric yields\footnote{The minus in the $T_{\tau\tau}$ component makes the fluid energy density positive in Lorentzian signature.}
\begin{equation}\label{ELEstreso}
T_{ab}=-\frac{M}{8\pi \ell^2}\, \left(\begin{array}{ccc}
 {2 } & 0 & {-8  n \sin ^2\frac{\theta }{2}} \\
 0 & - \ell^2 & 0 \\
- 8  n \sin ^2\frac{\theta }{2} & 0 & {32 n^2 \sin^4 \frac{\theta
   }{2}}\, - \ell^2\, \sin ^2\theta  \\
\end{array}
\right),
\end{equation}
where $a,b={\tau ,\theta,\phi }$. The boundary normalized Hamiltonian Killing vector, interpreted in fluid/gravity as the fluid velocity, is $ u= \partial_{\tau }$; in Euclidean signature this vector is spacelike. This ensures that in the flat limit this vector becomes null. On the other hand, the Cotton tensor evaluated on (\ref{ElTaubNUTgeneralrewritten}) reads
\begin{equation}\label{ELCotton}
C_{ab}=-{\frac{n}{\ell^4}\left(1-\frac{4n^2}{\ell^2}\right) }\, \left(\begin{array}{ccc}
 {2 } & 0 & {-8  n \sin ^2\frac{\theta }{2}} \\
 0 & - \ell^2 & 0 \\
 {-8  n \sin ^2\frac{\theta }{2}} & 0 & {32 n^2 \sin^4 \frac{\theta
   }{2}}\, - \ell^2
  \, \sin ^2\theta  \\
\end{array}
\right).
\end{equation}
The Cotton energy density is then computed to be $c=\ell^4  C_{ab}u^a u^b=- 2n (1-4 {n^2/ \ell^2}) $. We see here also that something special happens at the special points $n=\pm \ell /2$, for which the Cotton tensor vanishes as for $n=0$. 

Expressions (\ref{ELEstreso}) and (\ref{ELCotton}) satisfy the simple relation 
\begin{equation}
    T_{ab} = \frac{M\ell^4}{8\pi n(\ell^2-4n^2)} \, C_{ab}\,,
\end{equation}
which is manifestly compatible with the stress-tensor / Cotton tensor duality referred to above. The particular solution (\ref{ElTaubNUTgeneral}) that is self-dual corresponds to the Taub-NUT-AdS spacetime, $MG= n (1-4{n^2/ \ell^2})$, and, in the Euclidean theory, the complete boundary tensor evaluated on this solution yields 
\begin{equation}
T^+_{ab} \,\big|_{\text{NUT}}=0,\label{T0}
\end{equation}
and so vanishing energy (respectively, $T^-_{ab}\,|_{\text{NUT}}=0$ for the anti-self-dual solution with NUT charge $-n$). The vanishing of (\ref{T0}) is consistent with the fact that the mass (\ref{LAMasa}) is measured with respect to NUT spacetime as a reference background. The charge computation in terms of the Brown-York stress tensor would yields zero results if $MG= n (1-4{n^2/ \ell^2})$ is satisfied.

Going back to the question of parity (a)symmetry, we observe that expressions (\ref{ELEstreso}) and (\ref{ELCotton}) realize the fact that, under parity transformation, we get the map $T^{\pm }_{ab}\to T^{\mp}_{ab}$. In particular, $T^-_{ab}=0$ for $MG= n (4{n^2/ \ell^2}-1)$. The behavior of $T_{ab}$ and $C_{ab}$ under parity can be immediately observed from the fact that, while the component $T_{\tau \phi }$ is linear in $n$, the component $C_{\tau \phi }$ is proportional to a polynomial of even powers of $n$. This is related to the fact that the change $(\tau , \phi ,n) \to  (+\tau , -\phi ,-n)$ realizes the parity transformation; it leaves invariant $T_{ab}$ while changes the sign of $C_{ab}$. The same argument holds for the transformations $(\tau , \phi , n) \to  (-\tau , -\phi ,+n)$ and $(\tau , \phi ,n) \to  (-\tau , +\phi ,-n)$. In the fluid interpretation of the boundary theory \cite{Caldarelli:2012cm, Kalamakis:2020aaj}, the NUT charge is a monopolar source of vorticity. This is reminiscent of a magnetic monopole. Thence, the stress tensor / Cotton tensor duality is the fluid analog of electric / magnetic duality. Therefore, once again, the presence of $n$ in the boundary metric makes natural to consider both pieces in (\ref{Tpm}). 
\vspace{-0.5cm}

\section{Conclusions}\label{IV}

Let us summarize our results: in this paper, we have considered the method of topological renormalization in AlAdS spaces. The latter consists in adding to the gravitational action a topological term that, while suffices to render the Euclidean action finite and the variational principle well posed, provides a natural definition of the renormalized Noether charges. We used this prescription to work out the thermodynamics of AlAdS spacetimes, focusing on the physical properties of the Misner strings of both the Taub-NUT-AdS and Taub-Bolt-AdS solutions. This enabled us to correctly compute the contribution of the Misner string to the gravitational entropy by treating on the same footing the AdS and AlAdS sectors. We also investigated the effects of introducing the Chern-Pontryagin topological invariant in the gravity action, namely a parity odd dimension-four operator whose role is setting the Taub-NUT-AdS geometry as the reference background. We discussed how the presence of topological terms in the gravitational action contributes to the holographic stress tensor upon a suitable asymptotic expansion of the fields. While the relation between topological and holographic renormalization had previously been studied, our discussion here successes in treating the AdS and AlAdS sectors in equal footing as well as including the parity odd contributions. These contributions are responsible for the appearance of the Cotton piece in the boundary total stress tensor. This augmented stress tensor then identically vanishes on (anti-)self-dual backgrounds. 

The method employed here is generally covariant and could be adapted to other black hole solutions. In particular, it would be worth pursuing it for other AlAdS solutions, like accelerating black holes (AdS $C$-metric) or even more general Robinson-Trautmann spacetimes, whose thermodynamics recently received  renewed interest \cite{Appels:2016uha, Anabalon:2018qfv}. We plan to go back to this problem in future work with the intention to studying further the role of defects: the presence of the Misner string can be seen as a boundary defect. Studying this could allow us to make contact with the intertwining between renormalized volume and area in asymptotically hyperbolic spaces \cite{Anastasiou:2018mfk}. Furthermore, the defect corresponds to introducing corners in the boundary, so it would be interesting to connect our work with recent discussions on corners \cite{Freidel:2020xyx, Freidel:2020svx, Freidel:2020ayo}. Here, we focused on the entropy; however, the Noether-Wald's formalism is well-suited to study general asymptotic charges. Asymptotic charges have recently been studied for Taub-NUT/Bolt in locally flat spacetime in relation to the magnetic counterpart of BMS charges, see e.g.  \cite{Godazgar:2018qpq, Kol:2020ucd}. This raises the interesting question of the Ricci flat limit of our results, in the spirit of \cite{Ciambelli:2018wre, Compere:2019bua}. Finally, while the total boundary stress tensor $T^\pm_{ab}$ is complex in Lorentzian signature, in Euclidean signature it is real. Thus, unveiling its microscopic properties in the boundary CFT is challenging but in principle possible. In particular, we look forward to a more comprehensive understanding of the microscopic structure of the boundary dual of Taub-NUT/Bolt-AdS spaces, and its behavior under parity.
\vspace*{-0.5cm}

\begin{acknowledgments}
We thank D.~Flores-Alfonso, R. G.~Leigh, J.~Oliva and A. C.~Petkou for valuable exchanges and discussions. L. C. is supported by the ERC Advanced Grant \textsl{High-Spin-Grav}. The work of C. C. is supported by FONDECYT N$^o$~11200025.  J. F. acknowledges the financial support of ANID through the Fellowship 22191705. The work of G. G. is supported by CONICET grant PIP 1109-2017. The work of R. O. is supported by FONDECYT N$^o$~1170765.
\end{acknowledgments}

\bibliography{References}

\end{document}